\documentstyle[preprint,aps,graphicx]{revtex}

\title{Collective Chaos}
\author{Tatsuo Shibata, Kunihiko Kaneko}
\address{%
Department of Pure and Applied Sciences, 
University of Tokyo,
Komaba, Meguro-ku, Tokyo 153, Japan}
\date{\today}

\begin{document}
\draft
\maketitle

\begin{abstract}
An algorithm to characterize collective motion is presented,
with the introduction of ``collective Lyapunov exponent'',
as the orbital instability at a macroscopic level.
By applying the algorithm to a globally coupled map, existence of
low-dimensional collective chaos is confirmed,
where the scale of (high-dimensional)
microscopic chaos is separated from the
macroscopic motion, and the scale approaches zero in the thermodynamic limit.
\end{abstract}

\pacs{05.45+b,05.90+m,64.60.Cn}

Low-dimensional chaotic motion often  
arises from a system with many degrees of freedom.
A classical example is chaos in a fluid system 
(such as Rayleigh-B\'enard convection), where
very high-dimensional chaotic motion should underlie
at a molecular scale.  A canonical answer for the
condition to have low-dimensional chaos at a
macroscopic level is given by separation of
scales distinguishable from a microscopic level.
Still it is not clear how such separation is possible,
since chaos can lead to the amplification of a small-scale error.

To address the question, we note that
a certain coupled dynamical
system\cite{collective,Pikovsky1994,KK1995,Shibata1997,CML&CA,Oscillator} 
shows some lower dimensional collective motion for macroscopic variables, 
while microscopic variables keep high dimensional chaos.
To characterize such collective motion,
Lyapunov exponent at a macroscopic scale will be introduced,
which specifies the growth rate of error at macroscopic variables.
By studying the dependence of the exponent on the
length scale and the system size, it is shown how
the ``collective chaos'' is compatible with microscopic chaos,
and how they are separated  at the `thermodynamic limit'.

Here we adopt a `heterogeneous' globally coupled map (GCM) 
with a distributed parameter:
\begin{equation} 
x_{n+1}(i) = (1-\epsilon)f_i(x_n(i))
+\frac{\epsilon}{N}\sum_{j=1}^{N}f_j(x_n(j)) 
\label{eq:GCM}
\end{equation}
where $x_{n}(i)$ is the variable of the $i$'th element
($i=1,2,3,\cdots,N$) at discrete time $n$, and $f_{i}(x(i))$ is the
internal dynamics for each element.  For the dynamics we choose the
logistic map $f_i(x) = 1-a(i) x^2,$ where the parameter $a(i)$ for the
nonlinearity is distributed between $[a_0 - \frac{\Delta a}{2},a_0 +
\frac{\Delta a}{2}]$ as $a(i)=a_0 +\frac {\Delta a (2i-N)}{2N}$.
As a macroscopic variable,
we adopt the mean field,
\begin{equation}
h_n={1\over N}\sum_{i=1}^Nf_i(x_n(i)).
\end{equation}
in which the collective motion is contained. On the other hand,
chaos of $x_n(i)$ is referred to `microscopic' here.

Conventional GCM with identical parameters is given by $\Delta a=0$,
whose study has revealed 
clustering, chaotic itinerancy, 
partial ordering, and so forth \cite{Kaneko1990a}.  In particular, study of collective dynamics has
gathered much attention\cite{collective,Pikovsky1994,KK1995,Shibata1997}.
When the coupling $\epsilon$ is small enough,  
oscillation of each element is mutually desynchronized, 
and the effective degrees of freedom increase in
proportion to the number of elements $N$.
Still, a macroscopic
variable is found to show some kind of ordered motion distinguishable from
noise, ranging from torus to high-dimensional chaos
\cite{collective,Pikovsky1994,KK1995,Shibata1997}.

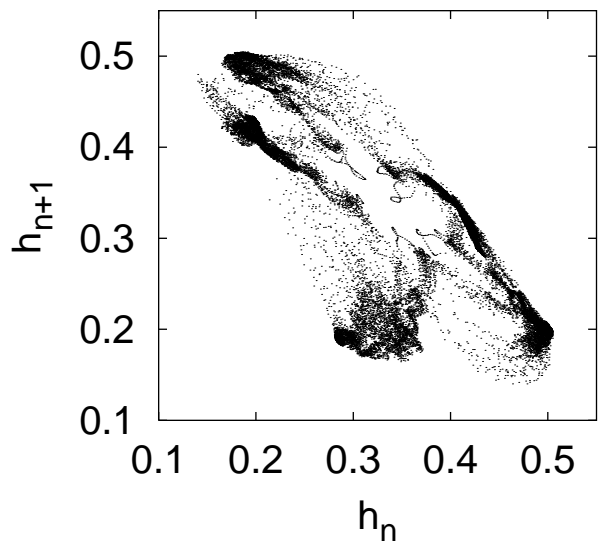
\begin{figure}
\caption{An example of return map for chaotic collective motion.
$a_0=1.92$, $\Delta a=0.088$, $\epsilon=0.1$, $N=10^7$.
Points $(h_n,h_{n+1})$ are plotted over $3\times10^4$ steps after
transient are discarded.}
\label{fig:rm}
\end{figure}

For instance, Fig.\ref{fig:rm} gives a return map of the mean field
dynamics of our model (\ref{eq:GCM}), which shows some pattern that may
suggest low-dimensional chaos.  Torus motion is also found by changing the
parameters \cite{Shibata1997}.  Here we try to characterize such
collective dynamics, and search for low-dimensional 
collective chaos.

First note that the conventional
Lyapunov exponents are not relevant to the characterization of collective 
motion.  Indeed, for small $\epsilon$, all the $N$ Lyapunov exponents 
of (1) are positive
(whose values are close to the exponent of a single logistic map
$x\rightarrow f_i(x)$).
No exponent corresponding to the mean field
motion is observed in this spectrum of $N$ exponents.
This apparent paradox can be resolved by noting the order of limit
to define the Lyapunov exponent.  In the calculation of the Lyapunov spectrum 
we take a 0 limit of disturbance applied to the orbit.  As long as we choose 
this limit first and then the thermodynamic limit($N \rightarrow \infty$), the 
$N$ Lyapunov exponents cannot characterize the collective motion \cite{torus}.
It is necessary to take the thermodynamic limit first and then the
limit of disturbance scale, to characterize the collective dynamics.

Since we are concerned with a system of large but finite $N$,
the above order of limit implies that we have to keep the disturbance amplitude finite.
To study such orbital instability, the finite-size Lyapunov exponent 
introduced by
Vulpiani and his coworkers\cite{Vulpiani} is useful.  It is given by
\begin{equation}
\lambda_{\delta_0}(\Delta)=\left\langle{1\over\tau}\log{\Delta\over\delta_0}\right\rangle,
\end{equation}
where $\tau$ is the maximum time such that $|x'_n-x_n|<\Delta$ for
trajectories $x_n$ and $x'_n$ starting from $x_0$ and
$x'_0=x_0+\delta_0$ respectively, while $\langle\cdot\rangle$ is an
average over the trajectories starting from different initial values.
The length scale $\Delta$ can be considered as the scale of
observation.

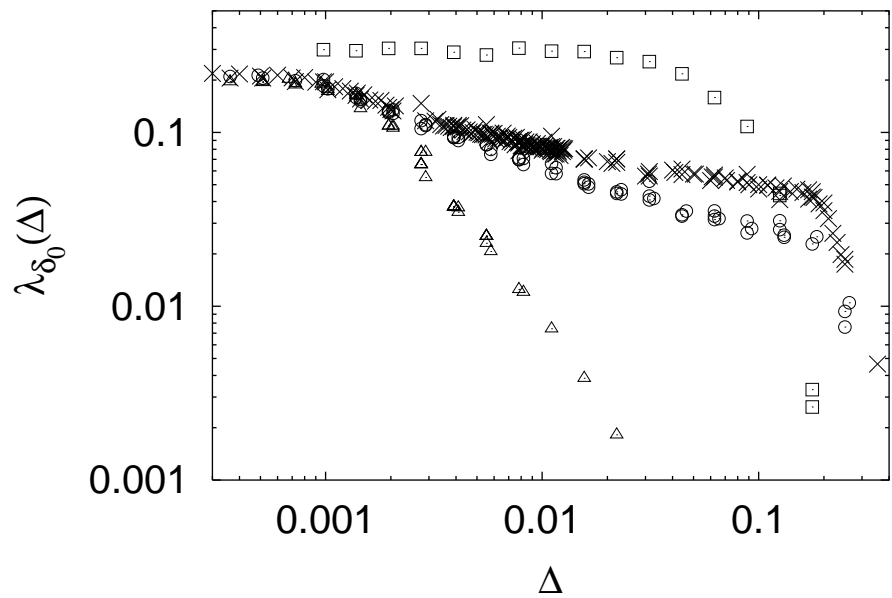
\begin{figure}
\caption{$\lambda_{\delta_0}(\Delta)$  is plotted for the model
(\protect\ref{eq:GCM}) 
with  $a_0=1.92$, $\Delta a=0.088$, $\epsilon=0.1$($\odot$), 
$a_0=1.9$, $\Delta a=0.05$, $\epsilon=0.098$($\times$), 
$a_0=1.9$, $\Delta a=0.05$, $\epsilon=0.11$($\bigtriangleup$), 
$a_0=1.69755$, $\Delta a=0.0$, $\epsilon=0.008$($\Box$), with $N=10^7$.
Initial perturbation amplitude $\delta_0$ is fixed at $1.0\times10^{-7}$.
For computation, displacement $h'_0=h_0+\delta_0$ is created by
perturbing the orbit as 
$x'_0(i)=x_0(i)+ \delta_0\times\sigma$, where $\sigma$ 
is a random number in $[-1,1]$.
Each point is obtained by averaging over 100 samples.
Specific choice of this perturbation scheme is irrelevant to our
results, as long as the collective variable is perturbed.
Adopting the algorithm to be
presented, the collective motion is shown to be torus ($\bigtriangleup$),
low-dimensional chaos ($\odot$ and $\times$),
and high-dimensional chaos ($\Box$).}
\label{fig:fin.lya}
\end{figure}

We have measured  the finite-size Lyapunov exponent 
for the macroscopic variable $h_n$ of our GCM (\ref{eq:GCM}).  
Here we perturb the orbit to give rise to a change
from $h_0$ to $h'_0=h_0+\delta_0$ (see the caption of Fig.\ref{fig:fin.lya}
for detailed description).
In Fig.\ref{fig:fin.lya}, the finite-size Lyapunov exponent
is plotted with the change of $\Delta$.

As long as the system size is finite, 
this finite-size Lyapunov exponent reflects not only the macroscopic
motion but also the microscopic chaos.
Indeed the exponent in Fig.\ref{fig:fin.lya} changes with the scale $\Delta$
and no clear plateau (except for $\delta_0\rightarrow 0$)
is visible.  On the other hand, 
if low-dimensional  macroscopic dynamics has a 
characteristic time scale separated from the  microscopic dynamics,
it will be possible to extract
the growth rate of perturbation in the collective motion
from the finite-size Lyapunov exponent for the macroscopic variable.
To do so, we postulate the following assumptions
that are expected to hold if the collective dynamics is
low-dimensional chaos or on a torus.

First note that in the limit, $\Delta\rightarrow0$ and
$\delta_0\rightarrow0$, the finite-size Lyapunov exponent
$\lambda_{\delta_0}(\Delta)$ converges to maximum Lyapunov exponent
$\lambda_m$, which is determined by the conventional Lyapunov exponents
for the microscopic variables $x_n(i)$ directly.

Considering that the collective dynamics appears by
coarse-grained macroscopic variables, we  postulate
that there are length scales (in the phase space) $\Delta\in[\Delta_m, \Delta_C]$,
where the macroscopic variable is characterized by 
``collective Lyapunov exponent'' $\lambda_C$.
Below  $\Delta < \Delta_m$ the microscopic chaos dominates, while
the orbit is out of the attractor (at a macroscopic level)
for $\Delta > \Delta_C$. 
To have low-dimensional collective dynamics,
it is postulated that $\lambda_C$ is independent of $N$
(as long as it is large enough), and that
$\Delta_m$ should approach zero with $N \rightarrow \infty$ while
$\Delta_C$ remains finite.

Based on the above assumptions, we can derive an approximate
form of  the finite-size Lyapunov
exponent against the scale $\Delta$.  Let $\delta_n$ denote the
distance from the original trajectory at time step $n$.
For the scale $\Delta<\Delta_m$, 
$\delta_n$ increases proportionally with $\exp{(\lambda_mn)}$.
Hence $\tau(\Delta)={1\over\lambda_m}\log{\Delta\over\delta_0}$ follows,
independently of the collective dynamics. 

On the other hand, for the scale $\Delta_m<\Delta<\Delta_C$,
$\delta_n$ is given as
$\delta_n\propto\exp{(\lambda_C n)}$ for a chaotic case with $\lambda_C>0$, or
$\delta_n\propto n^{\kappa}$ for a torus case
with $\kappa$ as a certain constant.  Corresponding to each collective motion,
$\tau(\Delta)$ and the finite-size Lyapunov exponent 
$\lambda_{\delta_0}(\Delta)$ are given by

\begin{equation}
\tau(\Delta)=\left\{
\begin{array}{lr}
{1\over\lambda_C}\log{\Delta\over\Delta_m}%
+{1\over\lambda_m}\log{\Delta_m\over\delta_0}&
\mbox{(chaotic case)}\\
\left(\Delta\over\Delta_m\right)^{1\over\kappa}
+{1\over\lambda_m}\log{\Delta_m\over\delta_0}-1&
\mbox{(torus case)}
\end{array}
\right.,
\end{equation}
and
\begin{equation}
\lambda_{\delta_0}(\Delta)=\left\{
\begin{array}{cl}
{\lambda_m\lambda_C\log{\Delta\over\delta_0}
\over
\lambda_C\log{\Delta_m\over\delta_0}+\lambda_m\log{\Delta\over\Delta_m}}%
&(\mbox{chaotic case})\\
{\log{\Delta\over\delta_0}
\over
{1\over\lambda_m}\log{\Delta_m\over\delta_0}
+\left(\Delta\over\Delta_m\right)^{1\over\kappa}-1}&(\mbox{torus case})
\end{array}
\right..
\label{eq.fin.lya}
\end{equation}

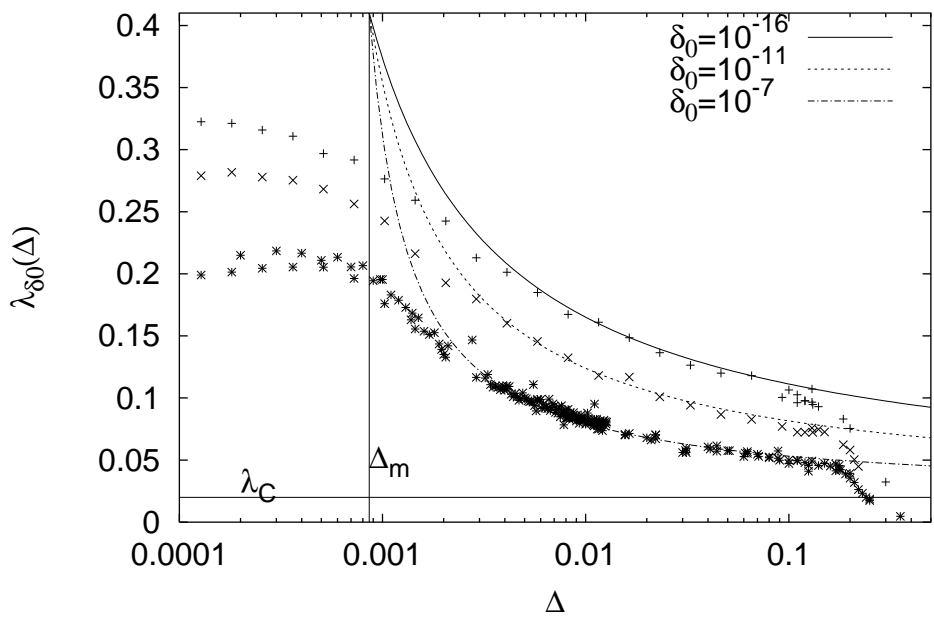
\begin{figure}
\caption{
The finite-size Lyapunov exponent $\lambda_{\delta_0}(\Delta)$ is
plotted as a function of $\Delta$ for several initial displacement
$\delta_0$.  The exponent is computed from the average over 
100 samples starting from different initial condition.  The curve from
Eq.(\protect\ref{eq.fin.lya}) fitted to the data is also
indicated. From the fitting,$\Delta_m=0.00086$ and  $\lambda_C=0.02$
are obtained, while $\lambda_m=0.41$ is
directly obtained from Eq.(\protect\ref{eq:GCM}).  $a_0=1.9$,
$\Delta a=0.05$, $\epsilon=0.098$, and $N=10^6$.
Note that although $\lambda_{\delta_0}(\Delta)$ for small $\Delta$ is 
slightly smaller than $\lambda_m$, it approaches $\lambda_m$
with $\delta_0\rightarrow0$.}
\label{fig:fitL}
\end{figure}

Fig.\ref{fig:fitL} shows
the behavior of $\lambda_{\delta_0}(\Delta)$ for our GCM,
with a fitting curve by Eq.(5). Here the parameters
$\lambda_C$ and $\Delta_m$ are obtained to fit the data for
several values of $\delta_0$ and $\Delta$.
The estimated exponent $\lambda_C$ is positive, and is
much smaller than the Lyapunov exponent $\lambda_m$ for microscopic chaos.
Thus there is a regime in which the mean field dynamics
shows low dimensional chaotic dynamics,
although there is no clear plateau corresponding to $\lambda_C$.
The value $\lambda_C$ is
smaller than $\lambda_{\delta_0}(\Delta)$.

It is then necessary to study $N$ dependence of $\Delta_m$, in order to
confirm the existence of the low-dimensional collective motion.
For this, it is convenient to transform Eq.(5) to
remove $\delta_0$ dependence of the data.
For it, we define $t(\Delta)$ as
$t(\Delta)=\tau(\Delta)+{1\over \lambda_m}\log{\delta_0}$,
which characterizes the time for amplification of error from
a certain scale independent of $\delta_0$.  From Eq.(4), we obtain
\begin{equation}
t(\Delta)=\left\{
\begin{array}{lr}
{1\over\lambda_C}\log{\Delta}%
+\left({1\over\lambda_m}-{1\over\lambda_C}\right)\log{\Delta_m}&
\mbox{(chaotic case)}\\
\left(\Delta\over\Delta_m\right)^{1\over\kappa}
+{1\over\lambda_m}\log{\Delta_m}-1&
\mbox{(torus case)}
\end{array}
\right..
\label{eq:t}
\end{equation}
Thus, the $N$ dependence of $\Delta_m$ appears as a shift of
constant in $t-\Delta$ plot, while 
$\lambda_C$ or $\kappa$ is given by a slope
in a suitable plot.

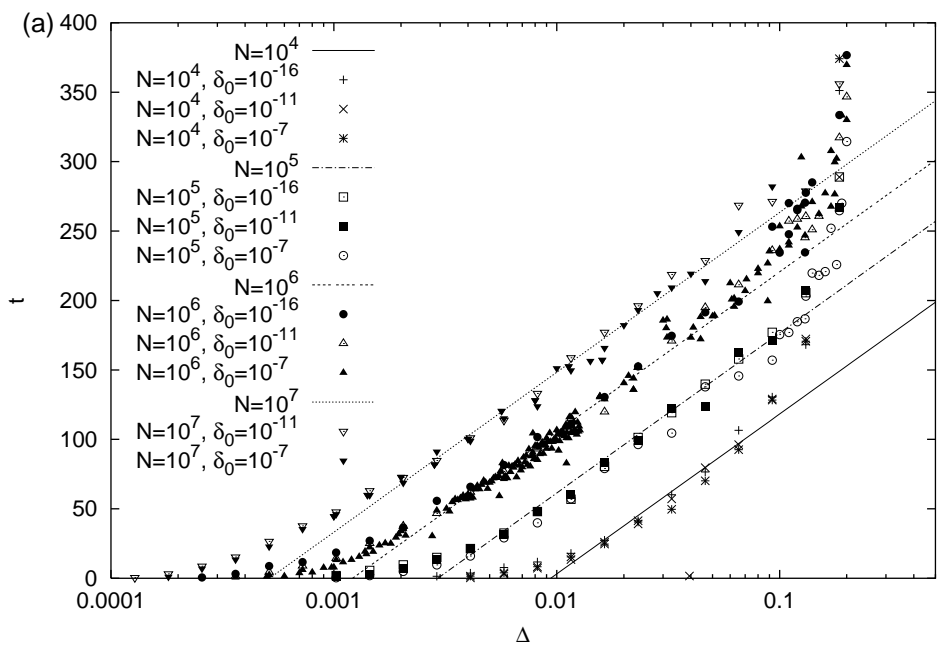
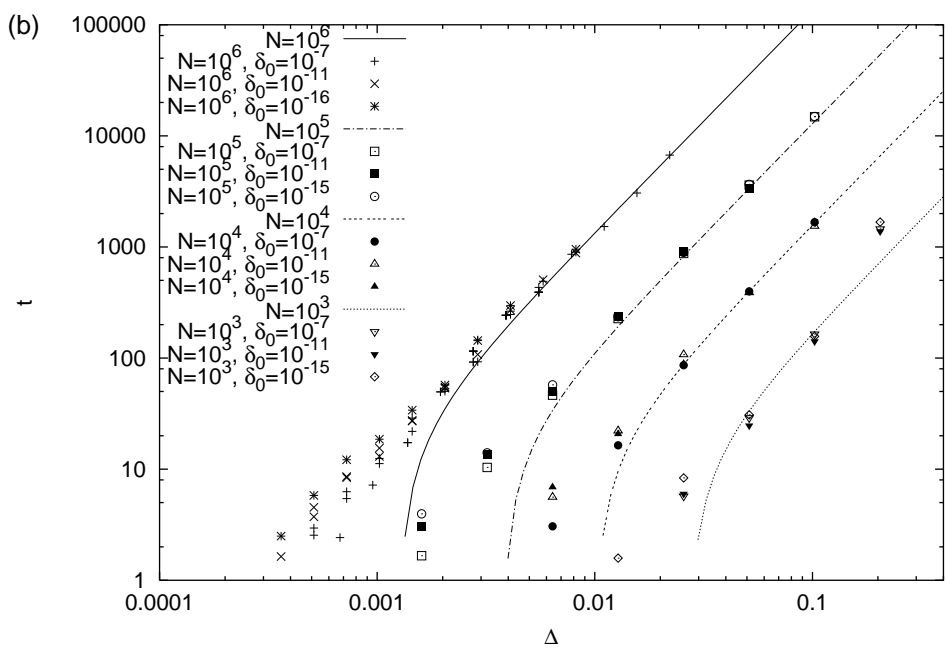
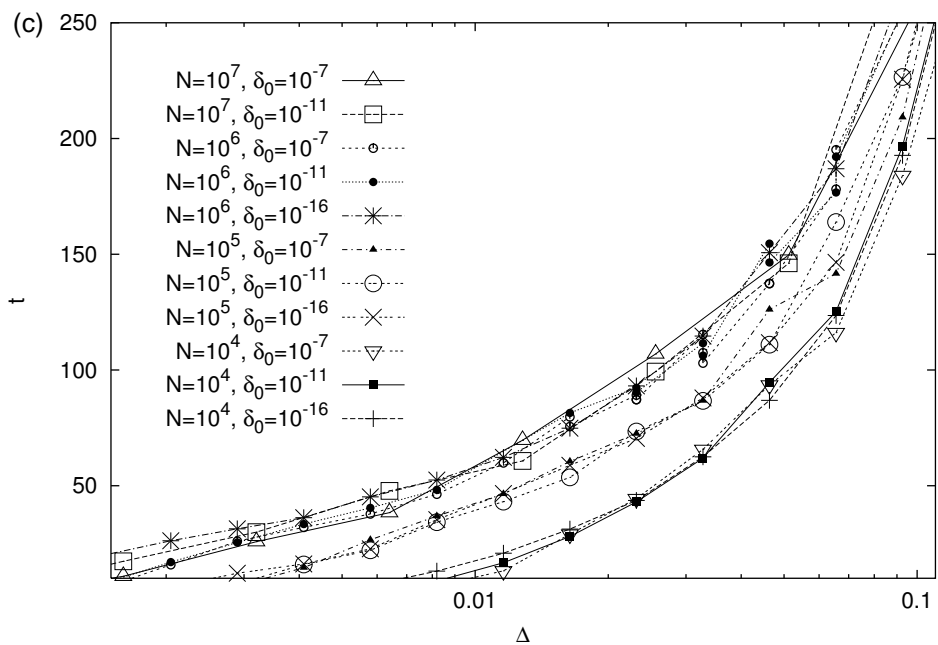
\begin{figure}
\caption{
The normalized time steps $t(\Delta)$ are 
plotted for $N=10^4$, $10^5$, $10^6$, and $10^7$,
with the fitted curves(\protect\ref{eq:t}).
(a) {\it chaotic case} (with a semi-log plot), for 
$a_0=1.9$, $\Delta a=0.05$, $\epsilon=0.098$.
(b) {\it torus case} (with a log-log plot), for
$a_0=1.9$, $\Delta a=0.05$, $\epsilon=0.11$.
The maximum Lyapunov exponent 
$\lambda_m=$0.41(a), 0.39(b) are obtained directly 
from our model(\protect\ref{eq:GCM}).
The parameters obtained by a least square fitting algorithm
give $\lambda_C=0.02.$(a),$\kappa=0.5$(b).
(c){\it high-dimensional case},
which does not obey Eq.(\protect\ref{eq:t}), 
(with a semi-log plot), for $a_0=1.6962$, $\Delta a=0$, $\epsilon=0.008$.
In this case, while the return map shows some structure,
$t$ for $N=10^6$ and $10^7$ are not separated any more.}
\label{fig:fitT}
\end{figure}

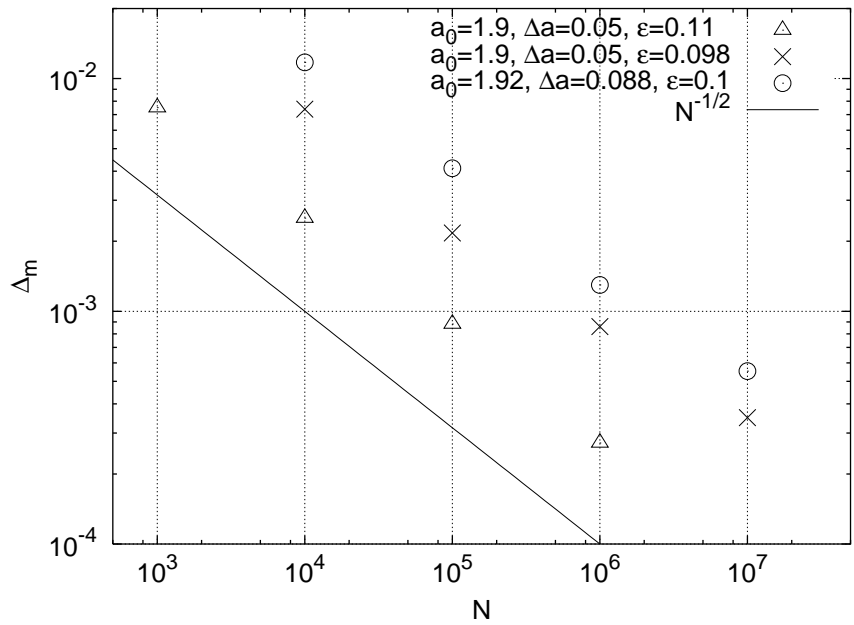
\begin{figure}
\caption{The microscopic length scales $\Delta_m$ are plotted as a
function of  $N$ for several parameters.
$\Delta_m$ is obtained from the fitting indicated in
Fig.\protect\ref{fig:fitT}.
The parameters are $a_0=1.92$, $\Delta a=0.088$,
$\epsilon=0.1$($\odot$, chaos), 
$a_0=1.9$, $\Delta a=0.05$, $\epsilon=0.098$($\times$, chaos), 
$a_0=1.9$, $\Delta a=0.05$, $\epsilon=0.11$($\bigtriangleup$, torus), }
\label{fig:tn}
\end{figure}

In Fig.\ref{fig:fitT}, $t$ is plotted as a function of $\Delta$.
As is shown in Fig.\ref{fig:fitT}(a), the slope of the semi-log plot
is independent of $N$.  The Lyapunov exponent $\lambda_C$,
characterizing the collective motion, is given by
the inverse of the slope, and is estimated as $0.02$.  
On the other hand, $\Delta_m$, given by the shift of the plots,
decreases with $N$, while $\Delta_C$ does not show significant change.
Thus the scale for the collective motion $\Delta_m < \Delta < \Delta_C$
increases with $N$. In
Fig.\ref{fig:tn}, $N$ dependence of $\Delta_m$ is plotted,
which gives $\Delta_m\sim {1\over\sqrt{N}}$,
whose form is expected from the central limit theorem.
Hence the emergence of low-dimensional collective chaos
at the thermodynamic limit is confirmed.

We have also applied the present algorithm to the case
with a collective torus motion.
Fig.\ref{fig:fitT}(b), ($t-\Delta$ plot), 
shows that $\kappa$, the inverse of the slope,
is 0.5, independent of $N$.  Indeed this exponent $1/2$ is
expected from the diffusion process of phase on the torus.
The decrease of $\Delta_m$ with $N$ is also plotted in Fig.\ref{fig:tn},
which again shows the expected decrease of $\Delta_m\sim {1\over\sqrt{N}}$.
Hence the collective torus motion is demonstrated.

In this letter,
we have proposed an algorithm to characterize the
collective (chaotic) motion, and applied to it to a GCM.
We have introduced collective Lyapunov exponent,
to characterize the growth rate of perturbation in the collective motion.
The microscopic chaotic motion exists at 
a small scale of the macroscopic variable, but
such scale $\Delta_m$ is shown to decrease as $1/\sqrt{N}$.
Hence, the macroscopic motion is separated from the microscopic motion
and the emergence of low-dimensional
collective motion with  $N\rightarrow \infty$
is confirmed.

Existence of low-dimensional collective chaos
in the presence of microscopic chaos has often been suspected\cite{Bohr}.
Indeed for a GCM with homogeneous elements (i.e., with  $\Delta a=0$), such 
low-dimensional collective chaos has not been observed so far.
In Fig.\ref{fig:fitT}(c), we have also applied our algorithm to this case.
The separation of scales is not clear and the data cannot be fitted
with (6). The shift of the plot gets smaller with the increase of $N$.
At leaset $\Delta_m$ does not decrease as  $1/\sqrt{N}$\cite{diff}.

Our $t-\Delta$ plot provides a tool to distinguish
low-dimensional collective chaos from high-dimensional one.
In the former case, the plot shifts as  $\log{(\sqrt{1/N})}$ with $N$,
while for the latter case such shift is not observed.
This distinction generally holds, even if the approximation to get
(6) may not be very good\cite{Vulpiani1998}.

Our estimation of macroscopic motion is realized 
for a system  subjected to microscopic chaos.
It is expected that our algorithm can be applied 
even if we do not know the equation of motion,
since the method of \cite{Vulpiani} is based on Wolf's algorithm\cite{Wolf}
developed for the estimate of Lyapunov exponents from experimental data. 
Thus, we hope that our method developed in this letter is applicable
to data obtained from experiments.

The authors thank the Supercomputer Center, Institute for Solid State
Physics, University of Tokyo for the facilities.
This work is partially supported by a Grant-in-Aid for Scientific
Research from the Ministry of Education, Science, and Culture
of Japan.


\begin{thebibliography}{99}

\bibitem{collective}
K. Kaneko, Phys. Rev. Lett. {\bf 65}, 1391 (1990);
Physica {\bf D55}, 368 (1992).
%
G. Perez and H. A. Cerdeira, 
Phys. Rev. {\bf A46}, 7492 (1992);
Physica  {\bf D63}, 341 (1993).
%
S. V. Ershov, A. B. Potapov, 
Physica {\bf D86}, 532 (1995);
Physica {\bf D106}, 9 (1997).
%
S. Morita, Phys. Lett. {\bf A211}, 258 (1996).
%
T. Chawanya, S. Morita, Physica, {\bf D116}, 44 (1998).
%
N. Nakagawa, T. Komatsu, Phys. Rev. {\bf E57}, 1570 (1998).

\bibitem{Pikovsky1994}
A. S. Pikovsky, J. Kurths, 
Phys. Rev. Lett. {\bf 72}, 1644 (1994);
Physica {\bf D76}, 411 (1994).

\bibitem{KK1995}
K. Kaneko, Physica {\bf D86}, (1995) 158.

\bibitem{Shibata1997}
T. Shibata, K. Kaneko, 
Europhysics Letters {\bf 38(6)}, 417 (1997);
preprint (1998) (xxx.lanl.gov chao-dyn/9802018,  submitted to Physica {\bf D}).

\bibitem{CML&CA}
H. Chat\'e, P. Manneville, 
Prog. Theor. Phys. {\bf 87}, 1 (1992).

\bibitem{Oscillator}
N. Nakagawa, K. Kuramoto,
Physica {\bf D80}, 307 (1995).
%
M.-L. Chabanol, V. Hakim, W.-J. Rappel,
Physica {\bf D103}, 273 (1997).

\bibitem{Kaneko1990a}
K. Kaneko, Physica {\bf D41},137 (1990). 

\bibitem{torus}
For example, all of the $N$ Lyapunov exponents are positive,
even if there appears quasiperiodic motion for the collective variable $h_n$
as $N$ goes to infinity \cite{Shibata1997}.

\bibitem{Vulpiani}
G. Paladin, M. Serva,  A. Vulpiani,
Phys. Rev. Lett. {\bf 74}, 66 (1995);
E. Aurell, et al., Phys. Rev. Lett. {\bf 77}, 1262 (1996).

\bibitem{Bohr}
T. Bohr, G. Grinstein, Y. He, C. Jayaprakash,
Phys.  Rev. Lett. {\bf 85}, 2155 (1987).
%
%

\bibitem{diff}
Hence, the scale of the macroscopic motion in GCM with $\Delta a=0$,
is not separated from the microscopic dynamics.
This gives a crucial difference between
the `heterogeneous' GCM and `identical' GCM.

\bibitem{Wolf}
A. Wolf, J. B. Swift, H. L. Swinney, J. A Vastano,
Physica {D16}, 285 (1985).

\bibitem{Vulpiani1998}
After the completion of the present manuscript, 
the authors are informed of the recent preprint
by M. Cencini, M. Falcioni, D. Vergni, A. Vulpiani
(xxx.lanl.gov chao-dyn/9804045),
where a related study to the collective
chaos is presented.

\end{thebibliography}
\end{document}